\documentstyle[preprint,aps]{revtex}
\begin{document}

\draft
\title{Impurity Scattering and Gap Structure in the Anisotropic Superconductor
UPt$_3$}

\author{Benoit Lussier, Brett Ellman, and Louis Taillefer}

\address{Department of Physics,
McGill University, 3600 University Street,
Montr\'eal, Qu\'ebec, Canada H3A 2T8
}

\date{\today}

\maketitle

\begin{abstract}

The thermal conductivity, $\kappa$,
of the heavy fermion superconductor UPt$_3$ was
measured down to T$_c$/10.
The absence
of a linear term in the temperature dependence as T$\rightarrow$0
strongly suggests there are no
zero energy quasiparticle excitations, in
contradiction with the gapless behaviour often inferred
from specific heat.
A non-vanishing anisotropy ratio $\kappa_c$/$\kappa_b$ as T$\rightarrow$0
establishes a new property of the gap structure: the presence of nodes along
the
c-axis.
Furthermore, recent calculations by Fledderjohann and Hirschfeld
show these cannot be point nodes with a linear k-dependence.

\end{abstract}


One of the central endeavours in the field of unconventional
superconductivity is the precise determination of the gap structure.
Numerous efforts are currently being made to establish whether or not a
d-wave gap is realized in the cuprate superconductors. In the heavy fermion
superconductor UPt$_3$,
a number of candidate
states are still possible
after 10 years of investigation.
Any state with a line node in the basal
plane of the hexagonal crystal structure, of which there are more than ten,
is compatible with most of the existing
data [1,2]. In particular, any such state is thought to be compatible
with the compelling anisotropy of transverse sound
attenuation [3].
A related issue of interest is the impact of impurity scattering on the low
temperature properties of such superconducting states. The appearance of
gapless behaviour, i.e. a residual density of quasiparticle excitations at
T=0, even for small concentrations of non-magnetic impurities
is under active investigation in the high T$_c$ cuprates.
In heavy fermions, no detailed study has been made of this issue -- even though
the specific heat of UPt$_3$, for example, has often been held as indicative of
a residual normal fluid in this compound --  nor have there been
quantitative
comparisons with existing calculations.

The detailed nodal structure of the gap and the effect of impurity
scattering will bear directly on the various scenarios for the
multicomponent phase diagram of UPt$_3$ [2]. Both properties
can only be studied at very low temperatures, typically in the
region of T$_c$/10. In a previous paper [4], we showed how the thermal
conductivity is a privileged probe of quasiparticles in UPt$_3$.
In particular, it is sensitive to
the anisotropy of the gap structure. Motivated by the
recent calculations of Fledderjohann and
Hirschfeld [5], we have now measured the thermal conductivity of UPt$_3$ down
to
T$_c$/10 and report two new facts: 1) there is no trace of a residual normal
fluid at T=0 (at the 1\% level) and 2)
the gap must go to zero along the
c-axis,
in addition to the well-established line node in
the basal plane.
A comparison with the results of Fledderjohann and Hirschfeld
further suggests that the gap
cannot vanish linearly at a point, thereby weighing against the
E$_{1g}$ representation for the superconducting order parameter of UPt$_3$
[2].

The details of the experiment and the crystal characteristics  are described
in Ref. [4]. The thermal conductivity $\kappa$(T) of UPt$_3$ was
measured for a heat current along each of the two high-symmetry directions of
the hexagonal lattice ({\bf J}//b-axis and {\bf J}//c-axis), as a function
of temperature down to T$_c$/10$\simeq$50 mK. The low-temperature results are
shown in Fig. 1, plotted as $\kappa$/T vs T, while the overall behaviour is
shown in the inset. The normal state behaviour below
T$_c^+$=0.5 K was obtained by
applying a magnetic field of 3 T $>$ H$_{c2}$(0) (dashed lines).
In this paper, we will only be concerned with the low temperature-low field
phase B, which exists below T$_c^-$=0.44 K [2].

In order to use heat conduction as a measure of electronic transport one
must
ensure that phonons do not contribute significantly, i.e.
that $\kappa_{ph}$$<<$$\kappa_e$ in the total conductivity
$\kappa$=$\kappa_{ph}$+$\kappa_e$. The safest estimate of the maximum
possible phonon contribution is obtained by using the formula
$\kappa_{ph}$=C$_{ph}$v$_{ph}$$\Lambda_{ph}$/3, where C$_{ph}$=$\beta$T$^3$
is the low
temperature phonon specific heat, v$_{ph}$ is the average sound velocity in
the direction of the heat current and $\Lambda_{ph}$ is the phonon mean free
path. One then assumes
that $\Lambda_{ph}$ takes on its maximum value, namely the size of
the crystal, equal to 0.7 mm [4]. From published data,
$\beta$=20 J K$^{-4}$ m$^{-3}$ [6], in agreement with the known sound
velocities [6]. The correct average velocity v$_{ph}$ [7] is
1880 m s$^{-1}$ for acoustic waves
travelling along the b-axis and 1440 for the c-axis [6]. The maximum heat
conduction by phonons in our crystal is therefore $\kappa_{ph}$=85 (67) T$^3$
mW K$^{-1} $cm$^{-1}$ for the b (c) axis. For T$<$150 mK, this represents
at most 15\% of the measured $\kappa_b$ and 6\% of $\kappa_c$.
Of course, scattering
by quasiparticles, which are present in sufficient numbers to account for
more than 85\% of $\kappa$,
will decrease $\Lambda_{ph}$ from this maximum possible value.
This decrease will occur
very rapidly as the temperature is increased, with $\Lambda_{ph}$
reaching a mere 5
$\mu$m at T=T$_c$ [4]. Therefore, $\kappa_{ph}$ is certainly less than the
upper bounds of 15\% and 6\%, and can thus be neglected.
We point out that while in our crystal the electronic mean free path is long
enough to ensure that $\kappa_{ph}$$<<$$\kappa_e$, this is less likely
to have been the case in
previous measurements on polycrystalline UPt$_3$, where $\kappa_e$ was 4 to
6 times smaller [1], not to mention measurements on other heavy
fermion compounds [1]. As a result, we believe
this is the first time heat transport by
heavy fermion quasiparticles is reliably measured down to T$_c$/10.
This allows us to examine the possibility of
a residual normal fluid at T=0 and it provides us
with a new and powerful probe of the gap structure.

Ideally, the question of a residual normal fluid should be answered by
low-temperature measurements of the specific heat. Unfortunately, this has
proven difficult both in UPt$_3$ and in the high T$_c$ cuprates,
for two reasons. First, a number of sizeable non-linear
contributions to C(T) exist at low temperature. In UPt$_3$, the
quasiparticle contribution, although large, is completely overwhelmed below
100 mK by a
huge upturn in C/T of ill-understood origin [8]. Secondly, even if a
residual linear term is extracted reliably, it cannot
automatically be attributed entirely to fermion excitations, for other
mechanisms also lead to a linear specific heat.
For UPt$_3$, the standard approach has been to extrapolate down from the
roughly linear behaviour of C/T observed above about 100 or 150 mK [1,2,8],
assuming it
to persist down to T=0. This
procedure yields an intercept at T=0, called $\gamma_0$, which typically
ranges from 10 to 40\% of the normal state value $\gamma_N$ in high quality
samples [1,8]. In our crystal, $\gamma_0$=16\% $\gamma_N$ [9].

In recent years, several theories have been proposed for this
extrapolated $\gamma_0$, taken to be an intrinsic property of the
quasiparticle fluid in UPt$_3$ [1,10,11]. In light of the strong sample
dependence, we are more inclined to regard this as an extrinsic property.
For unambiguous evidence about a possible residual normal fluid at T=0, we
therefore argue that one must
turn to a more reliable measurement, such as heat conduction, free from the
problems mentioned above. The observation of a finite $\kappa$/T at
T$\rightarrow$0, such as seen in the high T$_c$ compound
YBa$_2$Cu$_3$O$_7$, would be a direct
indication of zero-energy quasiparticle excitations.
It is clear from Fig.1 that a smooth extension of the $\kappa$/T
data to T=0 leads to a negligible intercept.
Of course, strictly speaking, an abrupt flattening off
of both curves below 40 mK cannot be excluded;
in this case the largest linear term would represent 2\%
(4\%) of its normal state value at T=0, equal to
L$_o$/$\rho_o$ [4], for {\bf J}//{\bf c} ({\bf J}//{\bf b}).
We therefore conclude that,
in the absence of any real sign of a residual
linear term in our data,
there is no evidence
for gapless excitations in UPt$_3$.

More quantitatively, we note that at our lowest temperature T=0.1 T$_c^-$,
$\kappa$/$\kappa_N$=3\% (on average), whereas an extrapolated C/T
for the same crystal gives
$\gamma$(T=0.1T$_c$)/$\gamma_N$=30\%.
Could it be that the mean free path due to impurity
scattering in the
superconducting state is 10 times smaller than in the normal state?
One of the few measurements of the quasiparticle
mean free path below T$_c$ comes from the
degradation of a heat current by vortices. The increase in the thermal
resistivity 1/$\kappa$ with
increasing magnetic field just above H$_{c1}$
is proportional to the density of vortices B/$\Phi_0$, the carrier mean free
path $l_0$ and the vortex effective diameter $D$. From their measurement of
this effect on a crystal of a quality similar to ours,
Behnia {\it et al.} [12] obtained $l_0D$ $\simeq$ 3000 nm$^2$ at T=0.2 K.
If $D$ is
taken to be the coherence length, then $D\simeq\xi(T=0.2 K)
\simeq\xi_0\simeq$ 10 nm [1,6], and $l_0 \simeq$ 300 nm. As we will see
below, this is very close to the mean free path we estimate
for the normal state.
Therefore, it appears unlikely that the relaxation time in the
superconducting state is suppressed by an order of magnitude.

The whole question of the impact of impurity scattering on unconventional
gap structures, and vice versa, was treated by several authors in the
mid-80s (see Refs.[1,13,14] and references therein).
Within a weak-coupling BCS theory,
Hirschfeld {\it et al.}
[13] showed that a self-consistent treatment of impurity scattering for a gap
with line nodes can lead to a residual density of quasiparticle states,
showing up as a finite C/T and $\kappa$/T at T=0.
Recent calculations of this kind were performed
by Fledderjohann and Hirschfeld [5]
for three uniaxial gap structures, each with a line node in the basal plane:
1) a polar gap (with no other nodes) and 2) two hybrid gaps (with in addition a
point node at each pole, i.e. along the c-axis). One of the hybrid gaps,
which we call
hybrid-I, vanishes linearly in k at the point nodes, while the
other, called hybrid-II, vanishes quadratically. Some of the states allowed by
hexagonal symmetry to which these
gaps correspond are listed in Table I.
The calculations so far assume a single ellipsoidal Fermi surface and s-wave
scattering, and they require two input parameters, the impurity scattering rate
$\Gamma_0$ and the scattering phase shift $\delta_0$ [5].
Much information about the nature of the scattering
can be obtained from the normal state properties.
 From the fact that the anisotropy of transport (both heat and charge) is
independent of temperature, i.e. that $\kappa_c$/$\kappa_b$=2.8 and
$\sigma_c$/$\sigma_b$=2.7 all the way from T=0.1 K (in the normal state) where
elastic scattering dominates to T=0.8 K where inelastic electron-electron
scattering dominates [4], it is very likely that
the anisotropy of 2.7-2.8 is due
to the Fermi velocity, and both impurity-electron collisions and
electron-electron collisions should be well-described by isotropic (s-wave)
scattering [5].
 From de Haas-van Alphen measurements (dHvA) [15],
the Fermi surface
is known to be made of several sheets, and a single ellipsoid is
certainly an oversimplification.
However, the usual Dingle plot
analysis yields fairly uniform scattering rates, with $\tau$ =
1/$\Gamma_{dHvA}$
= 2 - 4 x 10$^{-11}$ sec,
in crystals of a quality comparable to ours [15].
In temperature, this corresponds to $\Gamma_{dHvA}$ = 0.2 - 0.4 T$_c$.
The scattering rate $\Gamma_0$
appropriate for transport, which is less sensitive to small-angle
scattering and dephasing, will be smaller than $\Gamma_{dHvA}$
by a factor which depends on the
type of scattering. A value of $\Gamma_0$=0.1 T$_c$ seems quite reasonable,
corresponding to $l_0 = v_F / \Gamma_0$ = 400 nm [15].
A separate estimate, obtained from the shear viscosity at T$_c$ [14],
gives
$\Gamma$(T$_c$)=$\Gamma_0 (\rho(T_c) / \rho_0$) =
0.22 T$_c$, so that again $\Gamma_0$
 $\simeq$ 0.1 T$_c$.
Finally, Fledderjohann and Hirschfeld took
$\delta_0$=$\pi$/2, the unitary limit, seeing as
weaker scattering leads to sharp disagreement with observed properties [5].
Theoretical arguments for such a limit can be found in
Refs. [13] and [14], and in references therein.

The results of the calculations for $\kappa_b$ [5],
for two values of $\Gamma_0$ (0.1 and 0.01
T$_c$) are reproduced in Fig.2, alongside our data, as a function of reduced
temperature T/T$_c$.
The data are plotted as
$\kappa$/$\kappa_N$=$\kappa$(H=0)/$\kappa$(H$>$H$_{c2}$), where
$\kappa_N$=T/(a+bT$^2$) [4] and T$_c$=T$_c^-$=0.44 K. Given that b/a=0.25
K$^{-2}$ [4], $\kappa_N$ deviates from linear behaviour (elastic
scattering) by only 7\% at T=0.3T$_c$. This means that inelastic scattering
can safely be neglected in the calculations below that temperature, shown
therefore as $\kappa$/T normalized to 1 at T$_c$.
Only the curves for a polar and a hybrid-II gap structures are shown; the
corresponding curves for the hybrid-I gap lie in between [5].
Inspection of both data and calculated curves reveals that
the rapid increase in $\kappa$/$\kappa_N$ with temperature at such
low T/T$_c$ -- an order of magnitude faster than in a conventional
superconductor [4] -- is well reproduced by the calculations; this is a
convincing confirmation
of a line node in the basal plane of the gap structure of
UPt$_3$ (phase B).
However, the data show less curvature than either of the curves with
$\Gamma_0$/T$_c$=0.1 and, indeed, will not smoothly extrapolate to any
significant intercept at T=0, such as expected from the theory. In this
sense, the observed behaviour is more compatible with calculations based on
a much smaller $\Gamma_0$, say 0.01 T$_c$, with a temperature
dependence close to that of the polar gap,
although with
a magnitude closer to the hybrid-II gap.
Until the real Fermi surface is used both in the calculations and in the
estimates of a transport $\Gamma_0$, related directly to the measured
$\rho_0$,
it is difficult to make firm conclusions from this comparison.
Nevertheless, it does seem as though an unreasonably low scattering rate is
needed to keep the number of zero energy quasiparticle excitations
obtained in the current self-consistent calculations at the low level
observed in the experiment.

The anisotropy of electronic heat conduction has long been known
to be a useful probe
of gap anisotropy, even in s-wave superconductors.
In Ref.[4] we showed how the ratio
$\kappa_c$/$\kappa_b$
is a direct probe of the
anisotropy of the gap in UPt$_3$, insofar as
it is constant
above T$_c$ and starts falling immediately below T$_c$ (see inset of Fig.3).
The fact that it decreases rather than increases suggests there
are more thermally
excited quasiparticles with velocities along the b-axis than along the
c-axis. This could either result from a finite gap being larger along c than
along b (anisotropic s-wave gap), from the presence of nodes in the gap
along the b-axis
in the absence of any along the c-axis (polar gap), or from
the presence of
nodes along both axes (hybrid gap) provided the nodal structure is such
that more quasiparticles have {\bf v}//{\bf b}. This, for example, would
exclude
an axial gap (with only point nodes along the c-axis)
[4]. Note, however, that the current discussion relies on
the assumption
that the scattering rate does not change with temperature in some
unexpected way below T$_c$. To be free from such ambiguity, the analysis
must be done in a regime where the strong electron-electron scattering is
not important, namely below 150 mK or so.
Moreover, it is in the limit of T$\rightarrow$0 that a measurement of
$\kappa_c$/$\kappa_b$ becomes particularly useful.
Indeed at T$<<$T$_c$,
the regions of the gap very close to the nodes dominate the thermal
properties and a measurement of heat conduction
can then shed light on the detailed structure of the gap in the
vicinity of both high-symmetry directions.
These are the main justifications
for the present low temperature study.

Our results for $\kappa_c$/$\kappa_b$
below 0.3 T$_c$ are shown in Fig.3, normalized at T$_c$=T$_c^-$ (i.e. divided
by
2.8, the constant value of the normal state anisotropy).
The striking finding is that {\it the ratio does not go to zero as
T$\rightarrow$0}.
This definitively excludes not only an axial gap, but also a polar gap.
Indeed, the latter will give
$\kappa_c$/$\kappa_b$$\rightarrow$0 as
T$\rightarrow$0,
as a result of the clear difference between
excitation of quasiparticles with {\bf v//c} (across a finite gap) and
with {\bf v//b} (in the vicinity of a line node).
We conclude that the gap of UPt$_3$-phase B must have nodes along the
c-axis. No previous experiment imposed such a requirement on the gap
structure of this material, except perhaps the V-shaped gap features
observed in recent studies of point-contact spectroscopy [16]. Indeed,
all other measured properties are so far compatible with a polar gap [1].

A comparison with the calculations of Fledderjohann and Hirschfeld [5]
allows us to go further and gain insight into the specific nodal
structure near k$_x$=k$_y$=0. Their results on $\kappa_c$/$\kappa_b$
for the three uniaxial gaps with line nodes (polar, hybrid-I and
hybrid-II) are shown in Fig.3. In the absence of gapless behaviour
(i.e. for $\Gamma_0$ = 0.01 T$_c$), they find that
$\kappa_c$/$\kappa_b$$\rightarrow$0 as
T$\rightarrow$0 not only for the polar gap, as expected, but also for the
hybrid-I gap. In essence, a gap vanishing at a point node
with linear k-dependence does not cause as many quasiparticles to be excited
thermally as a gap vanishing along a line.
Remarkably, these authors found this not to be
true for a point node with {\it quadratic} k-dependence, and
the hybrid-II gap yields a {\it finite} value for
$\kappa_c$/$\kappa_b$ as
T$\rightarrow$0. More specifically, such a gap on a spherical or ellipsoidal
Fermi surface leads to no change in the anisotropy of heat conduction with
respect to the normal state [5].
Note that
the perfect isotropy will not be preserved for
the real Fermi surface [5].

We conclude that the gap structure of phase B is unlikely to be of the
hybrid-I type, as has been widely believed over the past few years [2].
Instead, our low temperature results favour a gap of the hybrid-II type,
being the only uniaxial gap with the correct limiting behaviour of
$\kappa_c$/$\kappa_b$ (see Table I).
Of course, states with non-uniaxial
symmetry should also be considered in future calculations.
In the context of the 2D theory for the multicomponent phase diagram of
UPt$_3$ (see Refs. [2] and [17], and references therein), this
appears to disqualify the (1,i) state of the E$_{1g}$, E$_{2g}$ and E$_{1u}$
representations as possible candidates for phase B, and leave only the (1,i)
state of the E$_{2u}$ representation with {\bf d}-vector parallel to the
c-axis [17].

In summary, measurements of the thermal conductivity of UPt$_3$
down to T$_c$/10 have
shed light on two important aspects of unconventional superconductivity: the
possibility of a gapless behaviour and the nodal structure of the gap function.
In the absence of any detectable linear term in $\kappa$ as T$\rightarrow$0,
we find no evidence for an intrinsic gapless regime, such as
previously inferred from the specific heat. A comparison with calculations
based on resonant impurity scattering and reasonable assumptions for the
scattering strength and anisotropy leads to a significant
discrepancy in the limiting value of $\kappa_b$/T.
The unusual observation of a finite value for the anisotropy ratio
$\kappa_c$/$\kappa_b$ as
T$\rightarrow$0 leads to new information about the gap structure of UPt$_3$
in phase B: the gap vanishes along the c-axis, and it does so with a special
k-dependence, not compatible with a linear point node but probably so with a
quadratic point node. This places the most stringent constraints so far on the
possible superconducting states, and hence on the various theoretical models,
for UPt$_3$.

We are grateful to A. Fledderjohann and P.J. Hirschfeld
for extensive
discussions and for allowing us to reproduce their calculations here.
This work
was funded by NSERC of Canada and FCAR of Qu\'ebec. L.T.
acknowledges the support of the Canadian Institute for Advanced Research and
the A.P. Sloan Foundation.

\begin{figure}
\caption{Low-temperature thermal conductivity of UPt$_3$,
divided by temperature, for a heat current
along the c-axis
(open circles) and the b-axis (solid
circles). Inset: $\kappa$/T up to 0.8 K.
The normal state behaviour,
obtained by applying a field above H$_{c2}$, is also shown (dashed lines).
}
\label{fig1}
\end{figure}

\begin{figure}
\caption{Thermal conductivity along the b-axis below 0.3 T$_c$, where
T$_c$=T$_c^-$=0.44 K. Upper frame: data plotted as $\kappa$/$\kappa_N$,
where $\kappa_N$=T(a+bT$^2$)$^{-1}$ (see text). Lower frame:
$\kappa$/T calculated for two uniaxial gaps with line nodes
(polar and hybrid-II) and for two values of the
impurity scattering rate $\Gamma_0$, 0.1 T$_c$ (solid lines) and 0.01 T$_c$
(dashed lines) (after Ref.[5]).
}
\label{fig2}
\end{figure}

\begin{figure}
\caption{Low-temperature behaviour of the anisotropy ratio
$\kappa_c$/$\kappa_b$, plotted as a function of reduced
temperature and normalized to unity at T$_c$.
The data (points) are compared with calculations for three gap
structures with a line node in the basal plane, with $\Gamma_0$=0.01 T$_c$
(after Ref.[5]).
The dashed line is a linear fit to the data below 0.3 T$_c$.
Inset: unnormalized data up to 0.8 K.}
\label{fig3}
\end{figure}

\begin{table}
\label{tab:veryoff}
\caption{
The gap structure of uniaxial states allowed by hexagonal symmetry
(for strong spin-orbit coupling), and the limiting value of the anisotropy
ratio $\kappa_c$/$\kappa_b$ expected as T$\rightarrow$0,
in the absence of gapless
behaviour (see Ref. [5]).
The nodal structures include a gap going to zero at a point
along the c-axis, either with a linear (LP) or a quadratic (QP) k-dependence,
and along a line in the basal plane (line node). Only odd-parity states with
{\bf d}//{\bf c} are listed (see Ref. [17]).
}
\begin{center}
\begin{tabular}{cccc}
Gap & Nodal structure & Superconducting states & $\kappa_c/\kappa_b$ as
T$\rightarrow 0$ \\
\hline
Anisotropic s-wave & none & $A_{1g}$ ($\Delta_b<\Delta_c$) & 0 \\
Axial I & LP &
$E_{1u}\ (1,i)$ & $\infty$ \\
Axial II & QP & $E_{2g}\ (1,i)$ & $\infty$ \\
Polar & line node & $A_{1u}$ & 0 \\
Hybrid I & line node + LP & $E_{1g}\ (1,i)$ & 0 \\
Hybrid II & line node + QP &
$E_{2u}\ (1,i)$ & 1 \\
\end{tabular}
\end{center}
\end{table}

\end{document}